\documentclass[iop]{emulateapj}

\newcommand{\chandra}{{\it Chandra}}
\newcommand{\uss}{USS 1558$-$003}

\slugcomment{Submitted to ApJ} 

\shorttitle{AGN in the USS 1558$-$003 Protocluster}
\shortauthors{Macuga et al.}

\begin{document}

\title{The Fraction of Active Galactic Nuclei in the USS 1558$-$003 Protocluster at $z = 2.53$}

\author{Michael Macuga\altaffilmark{1}, 
Paul Martini\altaffilmark{1,2},  
Eric D. Miller\altaffilmark{3}, 
Mark Brodwin\altaffilmark{4}, 
Masao Hayashi\altaffilmark{5}, 
Tadayuki Kodama\altaffilmark{6}, 
Yusei Koyama\altaffilmark{7}, 
Roderik A. Overzier\altaffilmark{8} 
Rhythm Shimakawa\altaffilmark{7},
Ken-ichi Tadaki\altaffilmark{5}, 
Ichi Tanaka\altaffilmark{7} 
}

\altaffiltext{1}{Department of Astronomy, The Ohio State University, Columbus, OH 43210, USA, macuga.3@buckeyemail.osu.edu, martini.10@osu.edu}

\altaffiltext{2}{Center for Cosmology and Astroparticle Physics, The Ohio State University, Columbus, OH 43210, USA}

\altaffiltext{3}{Kavli Institute for Astrophysics and Space Resaerch, Massachusetts Institute of Technology, Cambridge, MA 02139, USA}

\altaffiltext{4}{Department of Physics and Astronomy, University of Missouri - Kansas City, Kansas City, MO 64110, USA} 

\altaffiltext{5}{National Astronomical Observatory of Japan, Mitaka, Tokyo, 181-8588, Japan} 

\altaffiltext{6}{Astronomical Institute, Tohoku University, Aoba-ku Sendai, 980-8578, Japan} 

\altaffiltext{7}{Subaru Telescope, National Observatory of Japan, Hilo, HI 96720, USA} 

\altaffiltext{8}{Observat\'orio Nacional, Rio de Janeiro, RJ, Brazil}

\begin{abstract}

The incidence of Active Galactic Nuclei (AGN) with local environment is a potentially valuable probe of the mechanisms that trigger and provide fuel for accretion onto supermassive black holes. While the correlation between AGN fraction and environment has been well-studied in the local universe, AGN fractions have been measured for relatively few dense environments at high redshift. In this paper we present a measurement of the X-ray AGN fraction in the \uss\ protocluster associated with the $z=2.53$ radio galaxy 4C-00.62. Our measurement is based on a 100ks \chandra\ observation, follow-up spectroscopy from the Multi-Object Double Spectrograph on the Large Binocular Telescope, and broad and narrow band photometry. These data are sensitive to AGN more luminous than $L_{X}>2\times10^{43}$ erg s$^{-1}$ in the rest-frame hard X-ray band (2-10 keV). We have  identified two X-ray AGN at the redshift of \uss, one of which is the radio galaxy. We have determined that $2.0^{+2.6}_{-1.3}$\% of the H$\alpha$ emitters in the protocluster are X-ray AGN. Unlike most other high-redshift cluster progenitors studied with similar techniques, \uss\ does not have a significantly higher fraction of AGN than field galaxies at similar redshifts. This lower AGN fraction is inconsistent with the expectation that the higher gas fractions at high redshift, combined with the high galaxy densities and modest relative velocities in protoclusters, should produce higher AGN fractions. 

\end{abstract}


\section{Introduction}

One of the longest-standing questions in Active Galactic Nuclei (AGN) research is why AGN are only present in a fraction of all galaxies. The two requirements to produce significant accretion onto supermassive black holes (SMBHs) are a supply of fuel and some mechanism to remove the fuel's angular momentum. Observations of the most luminous starburst galaxies have provided strong evidence that mergers between gas-rich galaxies produce quasars \citep{sanders88,veilleux09}, and substantial theoretical work supports this scenario \citep{barnes91,hopkins05}. Many high-resolution images of quasar host galaxies also show evidence of interactions and mergers \citep{canalizo01,silverman11,barrows17}. However, careful statistical studies do not find a significant excess of interaction signatures in luminous quasar hosts relative to a control sample \citep{villforth17}. Thus while there is ample evidence for mergers in QSO hosts, direct proof of the merger hypothesis has been elusive.

The question of AGN fueling is more unclear for progressively more common, lower-luminosity AGN. While low-luminosity AGN are found in galaxies with younger stellar populations than otherwise similar control samples \citep{terlevich90,kauffmann03}, numerous studies have not found a single mechanism that can explain why some galaxies host AGN and others are quiescent \citep[e.g.][]{fuentes88,mulchaey97,martini03}. This may be because all of the mechanisms that have been proposed to explain low-luminosity AGN fueling, such as large-scale bars, minor mergers, asymmetries in the gravitational potential, and turbulence \citep{simkin80,elmegreen98,hopkins11} all contribute to some extent \citep[see][for a review]{martini04}.

An alternate way to test these scenarios for AGN fueling is to use the incidence of AGN as a function of environment. In the local universe, galaxies in dense clusters tend to have less star formation and less cold gas. This is because physical processes such as ram pressure stripping, evaporation by the hot ISM, tidal effects, and gas starvation lead to lower cold gas fractions in the cluster environment \citep{gunn72,cowie77,larson80,farouki81}. In addition, the relative velocities of galaxies in clusters are usually too high for galaxies to form bound pairs and merge. Measurements of the incidence of AGN in clusters of galaxies have shown that high-luminosity AGN are substantially rarer than in the field \citep{kauffmann04,popesso06}, although this is not the case for lower-luminosity AGN \citep{best05,haggard10}. These studies have shown circumstantial evidence that the same physical processes that produce fewer luminous starbursts in clusters also lead to a lower incidence of luminous AGN. 

There is a lot of additional information in how the AGN fraction as a function of environment evolves with redshift. Measurements of the fraction of star forming galaxies in clusters at $z=1$ are up to an order of magnitude greater than in the local universe \citep{butcher78, haines09, atlee12}. The evolution of the AGN fraction in clusters is similarly rapid over this redshift range \citep{galametz09,martini09,tanaka13}, and substantially greater than the increase in field AGN \citep{bundy08}. Those measurements have shown that while luminous AGN are substantially rarer in clusters than the field in the local universe, the AGN fraction is approximately the same in clusters and the field by $1 < z < 1.5$, and there is some evidence that the AGN fraction is greater in dense environments at $z > 2$ \citep{martini13,alberts16}.

To date there have been two detailed studies of the AGN fraction in significant galaxy overdensities at $z > 2$. These are color-selected galaxies in the QSO HS 1700+643 (hereafter HS 1700) protocluster at $z=2.3$ by \citet{digby10}, and the $z=3.09$ protocluster in SSA22 \citep{lehmer09}. Both of these studies found a higher AGN fraction in the overdensity than in a comparable field galaxy sample at similar redshift. These results are interesting because they appear to broadly confirm the expectation that their higher galaxy space density, combined with relatively modest relative velocities, should be a favorable environment for the fueling of supermassive black holes. Other high-redshift overdensities have also shown evidence for a higher incidence of AGN, such as the vicinity of the $z=2.16$ radio galaxy MRC 1138-262 \citep{pentericci02}, the 2QZ Cluster 1004+00 (hereafter 2QZ) at $z=2.23$ by \citet{lehmer13}, and the $z=1.6$ protocluster Cl 0218.3-510 by \citet{krishnan17}. It is less clear if these overdensities are the progenitors of local galaxy clusters. For example, 2QZ is only a factor of two overdense \citep{lehmer13}. Modest overdensities at high redshift may not become clusters by the present day, and in some cases it is difficult to determine if the overdensity has the potential of growing into a local, massive cluster, especially if there are limited redshift data \citep[e.g.][]{overzier05,chiang13,overzier16}. 

\begin{table}[ht]
	\caption{X-ray Sources\label{tbl:X}} 
	\begin{tabular} { l l l l c }
		\tableline
		\tableline
		RA & DEC & Net Counts & $f_{0.5-7keV}$ & Cluster Member \\
		(1) & (2) & (3) & (4) & (5)  \\        \tableline
		\tableline
		240.322 & -0.479 & 617.102 & 7.195 (0.291) & Yes \\
		240.282 & -0.451 & 223.511 & 3.418 (0.230) & No \\
		240.235 & -0.503 & 227.189 & 2.824 (0.189) & No \\
		$\cdot\cdot\cdot$ & $\cdot\cdot\cdot$ & $\cdot\cdot\cdot$ & $\cdot\cdot\cdot$ & $\cdot\cdot\cdot$ \\        \tableline
	\end{tabular}
	\tablecomments{\chandra\ catalog of 126 X-ray sources. The columns are: (1)-(2) the right ascension and declination (J2000); (3) net photon counts; (4) X-ray flux and uncertainty in the observed 0.5-7 keV band in units of $10^{-14}$ erg s$^{-1}$ cm$^{-2}$; (5) confirmed cluster member. The full table will be available as online material when the paper is published.}
\end{table}

In this paper we present our measurement of the AGN fraction in the \uss\ protocluster at $z=2.53$, which is associated with the radio galaxy 4C-00.62. This overdensity is the most significant one that does not have previous \chandra\ observations. \uss\ was first identified by \citet{kajisawa06} and \citet{kodama07} as part of a near-infrared imaging survey of the fields of radio galaxies with the Subaru telescope. \uss\ was consequently first identified with a method that is relatively more sensitive to stellar mass than star formation. Subsequent, narrowband studies by \citet{hayashi12}, \citet{hayashi16}, and \cite{shimakawa18} with Subaru have found a total of 107 H$\alpha$-emitting galaxies (hereafter HAEs) around the radio galaxy, which supports the conclusion that this region is a cluster in formation. We present our new observations from \chandra\ and the LBT in \S2. In \S 3 we describe our measurement of the AGN fraction and compare it to a field sample. In \S 4 we compare these results to measurements of other overdensities at high redshift. We summarize our results in \S 5. Throughout this paper we adopt $\Omega_{m}=0.3$, $\Omega_{\Lambda}=0.7$, and $H_{0}=70$ km s$^{-1}$ Mpc$^{-1}$ to calculate the luminosities of the AGN. 

\section{Observations and Data Analysis}

\begin{figure*}[ht]
    \plotone{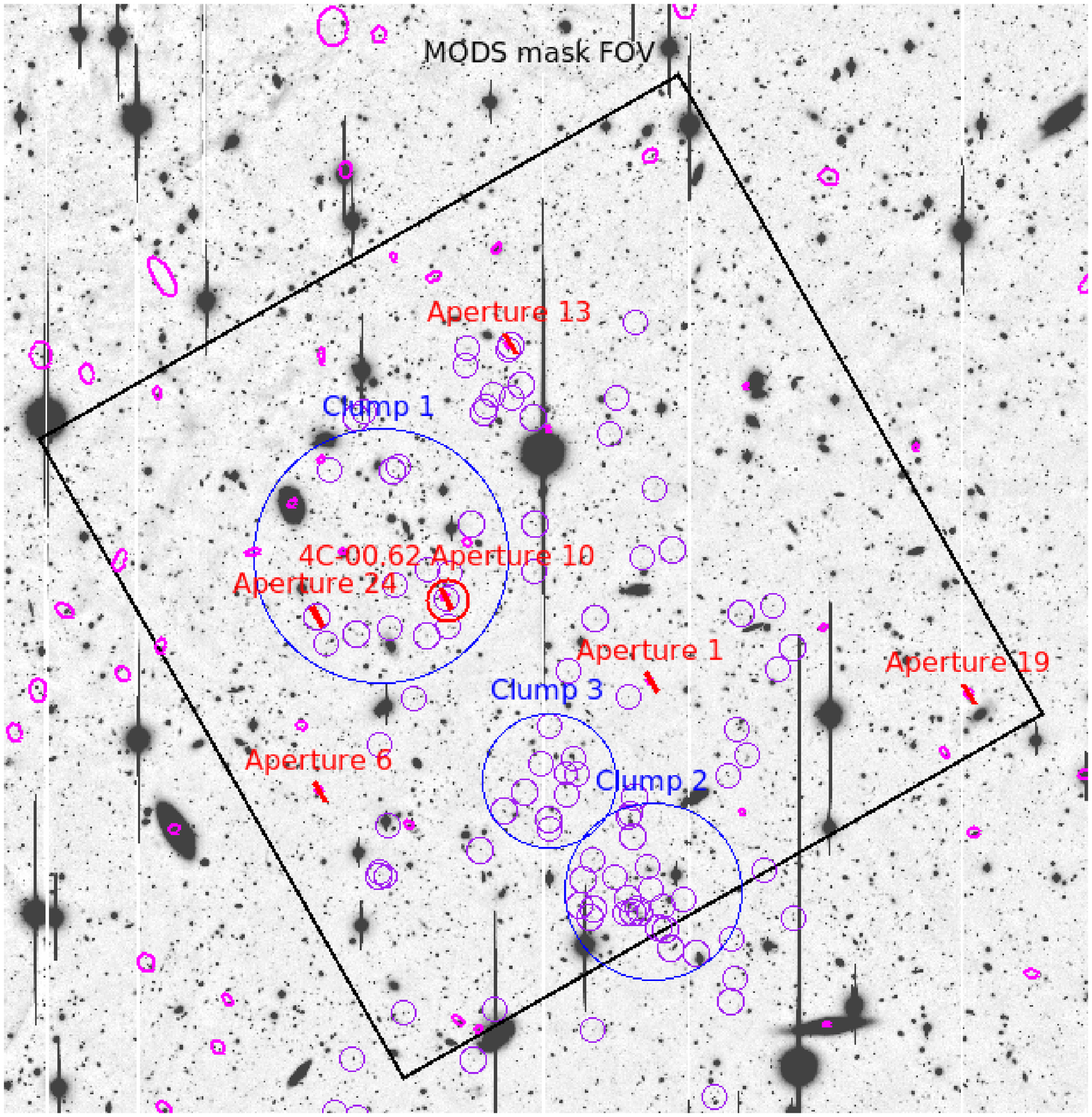}
    \caption{An $r-$band image from Subaru of the \uss\ protocluster. The large, black box shows the $6' \times 6'$ field of view of MODS. The large blue circles show three overdense clumps within \uss\ that were identified by \citet{hayashi12}.  The small purple circles are the HAEs from the \citet{hayashi16} catalog, and the small magenta ellipses are the X-ray sources from \chandra.  The red rectangles mark the five galaxies with confirmed spectroscopic redshifts from our LBT observations. The two members are the radio galaxy 4C-00.62 ({\it red circle}) and Aperture 013.\label{fig1}}
\end{figure*}

\subsection{Chandra}

We observed the \uss\ protocluster with the \chandra\ ACIS-I camera in June 2014. The X-ray images were split into two exposures of approximately 63ks (ObsID 16160) and 37ks (ObsID 16619) for a total integration of about 100ks. We processed the data with CIAO 4.6 and calibration products from CALDB 4.6.1. We merged the two datasets and identified 126 X-ray sources in the combined image with {\tt wavdetect} in a broad energy band from 0.5-7 keV. While the radio galaxy is clearly detected in the X-ray data, we did not detect any extended emission in its vicinity. 

We adopted a Galactic neutral Hydrogen column density of $N_H = 7.83 \times 10^{20}$ cm$^{-2}$ and a power-law index of $\Gamma = 1.7$ to calculate the fluxes of individual sources. Table~\ref{tbl:X} lists all of the X-ray sources. These data have sufficient depth to identify X-ray sources with $L_{X} = 2\times10^{43}$ erg s$^{-1}$ at $z=2.53$ in the rest-frame 2-10 keV band with $3\sigma$ significance. Point sources with this luminosity or greater are almost certainly AGN. 

\subsection{LBT}

We cross-matched the X-ray sources with visible and near-infrared Subaru images from \citet{hayashi12}. There are 27 X-ray sources within the $6' \times 6'$ field of view of the Multi-Object Double Spectrograph (MODS) at the Large Binocular Telescope (LBT). We assigned the highest priority to the counterparts of X-ray sources, followed by previously known cluster members, and then other sources with lower priority. Our mask design includes 22 of the 27 X-ray sources within the MODS field of view, and a total of 92 targets. We observed this slit mask for a total of 2.5 hours with MODS1 \citep{pogge10} in April and May of 2015. The image quality was slightly over $1''$ FWHM during two observations on 17 and 19 April and approximately $1''$ on 20 May. 

We were able to place a very high surface density of targets on the mask with the use of the MODS prism mode. This mode uses a ultra-low resolution $R = \lambda/\Delta\lambda \sim 200-500$ prism in each of the red and blue channels to disperse the light. The combined wavelength coverage of the two channels extends from 3300\AA\ to $1\mu$m. The two virtues of the prism mode are higher efficiency than the grating mode and the relatively short extent of the dispersed light on the detector, which allows multiple tiers of slits along the nominal dispersion direction. The prism mode has been used to successfully measure redshifts for emission-line galaxies as faint as $i=26$ mag \citep{adams15}. Figure~\ref{fig1} shows an overlay of the MODS field of view on an $r-$band image of the protocluster region from Subaru \citep{hayashi12}. The disadvantages of the prism mode are that the sky lines toward the red end are very blended, and the non-linear dispersion of the prisms increases the uncertainty in the wavelength calibration. 

We processed these data with a series of python scripts that correct for the bias level, fix bad pixels, and apply a flat field. We then used the MODS IDL pipeline in a similar manner to \citet{adams15} to apply the wavelength calibration, subtract the sky emission lines, extract the spectra, and apply a flux calibration. The very low resolution of these data complicates the wavelength calibration, particularly for slits near the edges of the mask that suffer from greater optical distortions. We therefore applied a zero-point shift to the wavelength calibration of all of the slits based on the locations of the 4165\AA{} \ion{Na}{1}  and 5577\AA{} \ion{O}{1} sky lines in the blue channel and the 5896\AA{} \ion{Na}{1} and 6300\AA{} \ion{O}{1} sky lines in the red channel. A simple zeropoint shift is not correct for a non-linear dispersion, and we used the locations of multiple sky lines to estimate the uncertainty in the wavelength calibration. 

\begin{figure*}[ht]
    \makebox[\textwidth]{\includegraphics[width=\paperwidth]{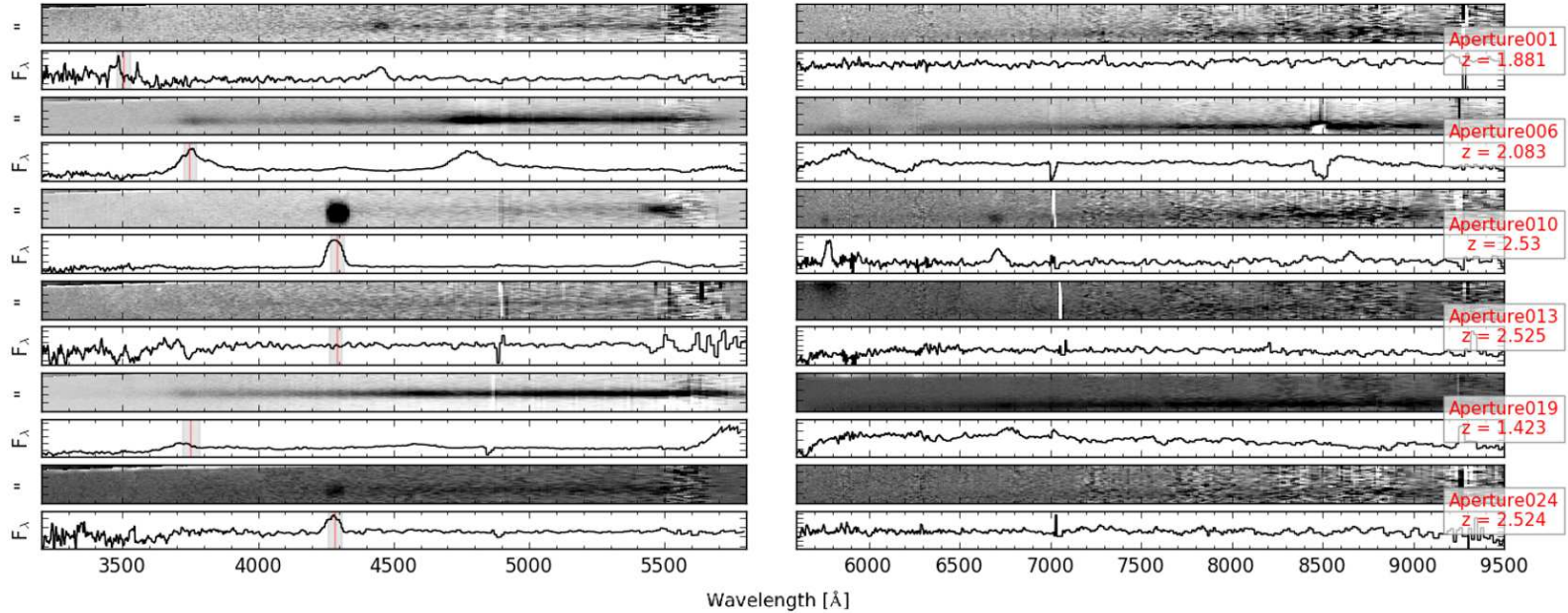}}
    \caption{Spectra for five galaxies with redshift measurements from MODS1 observations with the LBT. The left and right panels show the blue and red channels from MODS, respectively. For each galaxy, the upper panel shows the 2D spectrum and the lower panel shows the 1D spectrum. The red vertical line in the blue channel 1D spectra for Aperture 001, 006, 010, 013, and 024 shows the expected location of the Ly$\alpha$ emission at the protocluster redshift. This line for Aperture 019 represents a CIV detection.  The gray vertical bands surrounding the red lines represents the uncertainty in the wavelength calibration.\label{fig2}}
\end{figure*}

We measured redshifts for five galaxies in the field of the protocluster. Figure~\ref{fig2} shows the 1D and 2D spectra of these sources, and Table~\ref{tbl:z} lists additional data for these objects. All five objects have at least two spectral features that confirm the redshift. We used the velocity limits of $\pm 2000$ km/s from \citet{shimakawa14} to determine membership in the protocluster. These correspond to $2.5065<z<2.5535$.

Only one of the five X-ray sources with a secure redshift from our spectroscopy is consistent with protocluster membership. This source is the radio galaxy 4C-00.62 (Aperture 010), which motivated the original search by \citet{kajisawa06} and \citet{kodama07} for a protocluster at this location. The X-ray luminosity of 4C-00.62 is $L_X$ [2-10 keV] $= 2\times10^{45}$ erg s$^{-1}$.  We identified a second member (Aperture 024), although it is not an X-ray source. We cross-referenced the latest catalog of this region from \citet{hayashi16} and discovered one additional X-ray source at the protocluster redshift. While we had included this object in our MODS mask (Aperture 013),  we were unable to independently measure the redshift for this source. We adopt the redshift determination from \citet{hayashi16}, and the X-ray luminosity is $L_X$ [2-10 keV] $= 2.4\times10^{43}$ erg s$^{-1}$. Just one emission line is visible for three additional X-ray counterparts. While their redshifts are consequently ambiguous, the observed wavelengths of the emission lines are inconsistent with membership in the protocluster. We found no further members with a search of the NASA Extragalactic Database. We therefore base our analysis on the detection of two X-ray AGN in the \uss\ protocluster. 

\section{AGN Fractions}

\begin{table*}[ht]
	\begin{center}
	\caption{Redshift Measurements\label{tbl:z}} 
	\begin{tabular} { l l l c c c c c l l }
		\tableline
		\tableline
		Aperture & RA & DEC & X-ray & HAE & Member & Clump & z & $f_{0.5-7keV}$ & Line IDs \\
		(1) & (2) & (3) & (4) & (5) & (6) & (7) & (8) & (9) & (10)  \\        \tableline
		\tableline
		001 & 16:01:10.63 & -00:29:26.7 & Yes & No & No & N/A & 1.881 & 9.72 (1.36) & CIV, MgII \\
		006 & 16:01:21.40 & -00:30:21.4 & Yes & No & No & N/A & 2.083 & 23.2 (1.8) & Ly$\alpha$, CIV, CIII, MgII \\
		010 & 16:01:17.33 & -00:28:46.5 & Yes & Yes & Yes & 1 & 2.530 & 72.0 (2.9) & Ly$\alpha$, SiIV, CIV, CIII \\
		013 & 16:01:15.28 & -00:26:41.7 & Yes & Yes & Yes & N/A & 2.525 & 0.88 (0.35) &  \citet{hayashi16} \\
		019 & 16:01:00.23 & -00:29:32.0 & Yes & No & No & N/A & 1.423 & 8.9 (1.0) & CIV, CIII, MgII \\
		024 & 16:01:21.54 & -00:28:55.4 & No & Yes & Yes & 1 & 2.521 & $<0.88$ & Ly$\alpha$, LyB \\		\tableline
	\end{tabular}
	\tablecomments{Redshift measurements for galaxies in the field of the \uss\ protocluster. Three galaxies are members of \uss, but only two are X-ray sources. The columns are: (1)  Aperture ID from our MODS slit mask; (2)-(3) the right ascension and declination (J2000); (4)-(6) identification of the galaxy as an X-ray source, an H$\alpha$ emitter based on \citet{hayashi16}, and membership in the protocluster; (7) identification with the clumps defined in \citet{hayashi12}; (8) redshift; (9) X-ray flux and uncertainty in the observed 0.5-7 keV band in units of $10^{-15}$ erg s$^{-1}$ cm$^{-2}$; (10) features used to determine the redshift. The redshift for Aperture ID 013 is from \citet{hayashi16}. The redshift uncertainties are $\pm 0.001$ for Aperture ID 001, 006, 010, and 019 and $\pm 0.006$ for Aperture ID 024. LyB is the Lyman Break.}
	\end{center}
\end{table*}

\subsection{\uss\ AGN Fraction}

We have identified two X-ray sources in the \uss\ protocluster.  The denominator for the AGN fraction can be defined in a number of ways. We start with the catalog 100 HAEs from \citet{hayashi16}. These sources are based on an extraordinarily deep (9.7 hr) NB2315 narrowband image with the MOIRCS instrument on the Subaru telescope, $J$ and $K_s$ broadband images, and deep F160W images obtained with the WFC3 instrument on HST. There are also deep $B$, $r'$, and $z'$ images from Subaru that were previously presented in \citet{hayashi12}. The narrowband images have a $3\sigma$ flux limit of $f > 1.1 \times 10^{-17}$ erg s$^{-1}$ cm$^{-2}$, which corresponds to a luminosity of L(H$\alpha$ + \ion{N}{2}) $=5.8 \times 10^{41}$ erg s$^{-1}$ \citep{hayashi16}. \citet{shimakawa18} report that 49 of these HAEs have either spectroscopic redshifts or were also detected in redshifted Ly$\alpha$ narrow band images. The other 58 were selected by $r'JK_s$ and $Br'K_s$ colors. Both of our X-ray sources are in the HAE catalog, so we conclude that X-ray AGN with $L_X > 2 \times 10^{43}$ erg s$^{-1}$ constitute $2.0^{+2.6}_{-1.3}$\% of the HAEs with $L_{H\alpha} > 5.8 \times 10^{41}$ erg s$^{-1}$ in the protocluster. Note that throughout this paper we use double-sided 1$\sigma$ confidence limits from the methodology of \citet{gehrels86}. While there are 16 HAEs that fall outside of the MODS field of view, none of them are X-ray sources, so these objects do not affect our estimate of the AGN fraction. There are also no other HAEs with \chandra\ detections.

There are two potential sources of uncertainty in this estimate of the fraction of HAEs that are AGN. One is that we do not have spectra for all of the X-ray sources, and the other is that there could be other HAEs outside of the 39 arcmin$^2$ field surveyed with MOIRCS. Neither of these are significant sources of uncertainty. The first is not a source of uncertainty because the X-ray data completely overlap the MOIRCS images and consequently all of the HAEs. The second is almost certainly true, that is there are likely HAEs outside of the MOIRCS footprint. However, none of our confirmed X-ray sources are outside this footprint. As the MOIRCS field should be unbiased with respect to any AGN, our measurement should be representative of the AGN fraction of the protocluster. The one exception is that the field was selected to contain the radio galaxy 4C-00.62. If we remove this object, then the AGN fraction is $1.0^{+2.3}_{-0.8}$\%. 

Another important consideration is that the denominator is just based on HAEs. There are additional galaxies in the protocluster that are not emission-line objects, or at least fall below the sensitivity threshold of the narrowband data. While there are many galaxies in the field of the protocluster with colors consistent with membership, the spectroscopic redshift information for those galaxies is highly incomplete. We therefore only consider the AGN fraction in the HAE population in this paper. 

\subsection{Field AGN Fraction}

\begin{figure*}[ht]
    \plotone{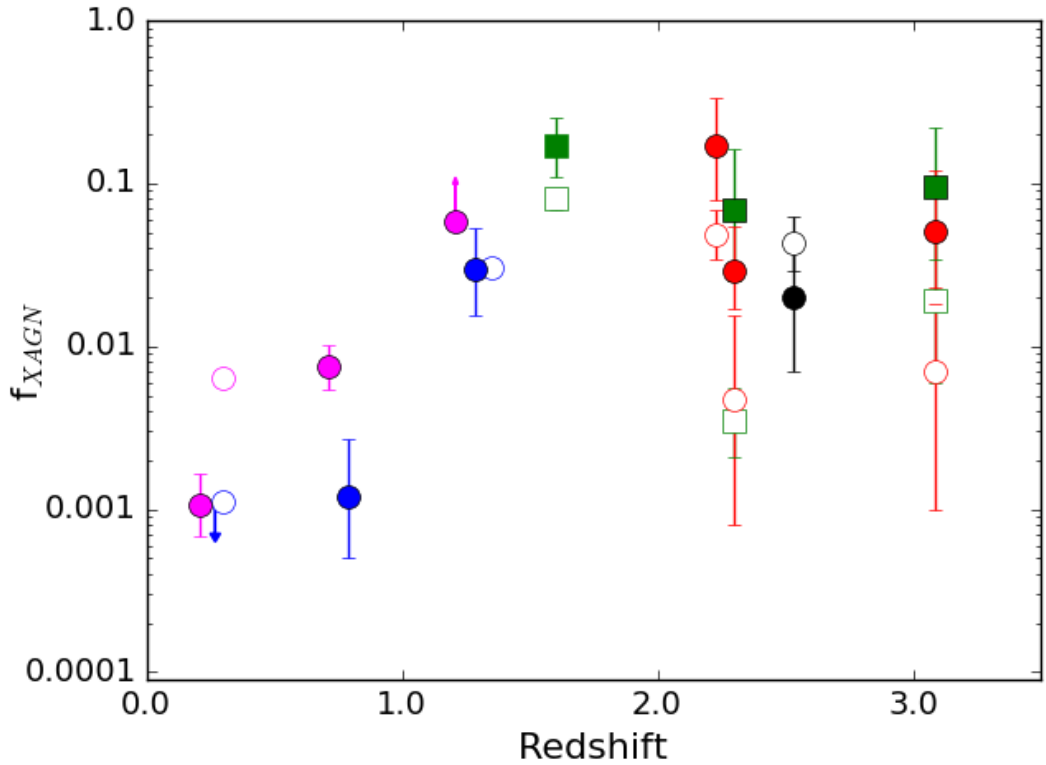}
    \caption{Evolution of the AGN fraction in clusters or protoclusters and the field. Cluster or protocluster measurements are solid symbols, and the field measurements are open symbols. The cluster AGN fraction is lower than the field fraction at $z<1$, comparable at $1<z<1.5$, and higher at $z>2$. \uss\ is the filled black circle at $z=2.53$, and the corresponding field fraction is the open black circle. Measurements selected with emission lines are shown as red circles, and measurements selected via broad-band colors are shown as green squares. The blue and magenta symbols at $z<1.5$ represent averages for large samples of clusters, not individual overdensities (see \S~\ref{sec:dis}). \label{fig:sum}}
\end{figure*}

Our X-ray and near-infrared broad and narrowband imaging data do not span a large enough field of view to measure the field AGN fraction at $z = 2.5$, as cluster progenitors can span $5-10$ Mpc at high redshift \citep{chiang13}. We therefore use data from the HiZELS project \citep{geach12,sobral13}. \citet{lehmer13} used these data to estimate the field AGN fraction and report there are 210 HAEs at $z = 2.23$ with $f_{H\alpha} > 5 \times 10^{-17}$ erg s$^{-1}$ cm$^{-2}$ (90\% completeness) that overlap the C-COSMOS X-ray data \citep{elvis09,puccetti09}. \citet{lehmer13} identified 10 \chandra\ sources in these 210 HAEs for a field AGN fraction of $4.8^{+2.0}_{-1.4}$\%, which is higher than but consistent with our cluster measurement at less than $2\sigma$. 

Both the X-ray and narrowband imaging for the C-COSMOS/HiZELS data have different sensitivity limits than our data for \uss. The HAE limit quoted by \citet{lehmer13} corresponds to a luminosity $L_{H\alpha} = 1.9 \times 10^{42}$ erg s$^{-1}$, or nearly a factor of four higher than the limit for the very deep MOIRCS data. Their X-ray luminosity threshold is $L_X > 10^{44}$ erg s$^{-1}$ in the 2-10 keV band, which is about a factor of five higher than our limit of $L_X > 2 \times 10^{43}$ erg s$^{-1}$. We do not attempt to correct for potential evolution of the X-ray and HAE fraction between $z=2.53$ and $z=2.23$ as this represents a mere 360 Myr. 

As both the H$\alpha$ and X-ray luminosity limits are higher for the field sample, we recalculate the \uss\ AGN fraction with these same limits. The higher X-ray luminosity threshold eliminates the AGN in Aperture ID 013, and only 40 HAEs are above the higher H$\alpha$ luminosity threshold. The net result is that the AGN fraction in \uss\ changes to $2.5^{+5.7}_{-2.1}$\%, and only the radio galaxy 4C-00.62 remains in the numerator. This is similar to our previous estimate of $2.0^{+2.6}_{-1.3}$\% to the H$\alpha$ and X-ray flux limits of our data, and remains smaller but consistent with the field AGN fraction. If we eliminate the radio galaxy 4C-00.62 because it motivated the initial identification of the protocluster, then we can only place an upper limit of 4.6\%  on the AGN fraction for the protocluster with these higher thresholds. 

\section{Discussion} \label{sec:dis} 

\begin{table*}[ht]
    \begin{center}
    \caption{AGN Fractions in Overdensities at High Redshift\label{tbl:sum}}
    \begin{tabular} { l c l l l l l l }
        \tableline
        \tableline
        Cluster         & z     & Selection & $L_{X,lim}$ & $N_{AGN}/N_{Tot}$ & $f_{AGN}$ & $f_{AGN}$ Field & Ref \\ 
        (1) & (2) & (3) & (4) & (5) & (6) & (7) & (8) \\
        \tableline
        \tableline
        Cl 0218.3-0510   & 1.6  & colors & $10^{42}$        & 8/46  & $17^{+8.6}_{-6.0}$\%      & $8^{+1.4}_{-1.2}$\%           & \citet{krishnan17} \\ 
        2QZ 1004+00      & 2.23 & HAE    & $10^{44}$        & 3/18  & $17^{+16.2}_{-9.1}$\%      & $4.8^{+2.0}_{-1.4}$\%     & \citet{lehmer13} \\ 
        QSO HS 1700+643  & 2.30 & BX/MD  & $5\times10^{43}$ & 2/29  & $6.9^{+9.2}_{-4.4}$\% & $0.35^{+0.21}_{-0.14}$\%  & \citet{digby10} \\ 
                         &      & LAE    &                  & 3/106 & $2.9^{+2.9}_{-1.6}$\% & $0.47^{+1.08}_{-0.39}$\%  & \\ 
        \uss\            & 2.53 & HAE    & $2\times10^{43}$ & 2/100 & $2.0^{+2.6}_{-1.3}$\% & $4.3^{+2.0}_{-1.4}$\%     & This study \\ 
        SSA 22           & 3.09 & LBG    & $3\times10^{43}$ & 2/21  & $9.5^{+12.7}_{-6.1}$\%& $1.9^{+2.6}_{-1.3}$\%     & \citet{lehmer09} \\ 
                         &      & LAE    &                  & 2/39  & $5.1^{+6.8}_{-3.3}$\% & $0.7^{+1.6}_{-0.6}$\%     & \\ 
        \tableline
    \end{tabular}
    \tablecomments{Summary of high-redshift overdensities with AGN fraction measurements. Col.(1)-(2): Name of the environment and the redshift; (3) Selection method to identify the parent sample of galaxies in the overdensity and field; (4) Limiting X-ray luminosity threshold for the study in erg/s; (5) number of AGN and total number of galaxies in the overdensity; (6) AGN fraction in the dense environment; (7) AGN fraction in the field selected with the same criteria; (8) Reference. There are two entries for HS 1700 and SSA22 because both studies used more than one selection method to identify AGN in the overdensity and the field.}
    \end{center}
\end{table*}

The AGN fraction in \uss\ and the field are consistent with one another, which is in sharp contrast with the higher AGN fraction measured for other overdensities at similar redshifts. The most similar study to ours is the \citet{lehmer13} analysis of the 2QZ structure at $z=2.23$, as it also used narrowband imaging to identify HAEs, although the overdensity is not as significant. They found seven X-ray sources in the 22 HAEs within the field of view of their \chandra\ data for an AGN fraction of $32^{+17}_{-12}$\%. However, they note that four of the seven were previously known AGN, and if those four are not included the AGN fraction is $17^{+16.2}_{-9.1}$\%. Both fractions are substantially higher than our measurement. If 17\% represents the true AGN fraction in HAEs at $z>2$, then we would have identified 18 in our 107 HAEs or 7 in the more luminous subsample. If the true fraction is 17\%, the binomial probability of observing two or fewer AGN in a sample of 107 is $< 10^{-6}$. For one or fewer AGN in a sample of 40 it is $<5 \times 10^{-3}$. \citet{lehmer13} also determined that the field AGN fraction is $4.8$\%, and therefore the AGN fraction in the 2QZ overdensity is $3.5$ times higher than the field fraction. This is the same field sample that we used for \uss.

The \citet{lehmer09} study of the SSA22 protocluster at $z=3.09$ measured that the AGN fraction is $9.5^{+12.7}_{-6.1}$\% for the Lyman Break Galaxies and $5.1^{+6.8}_{-3.3}$\% for the Lyman-$\alpha$ emitters. These values are both larger than we measure for \uss, although they use different selection methods and luminosity thresholds. \citet{lehmer09} also report that both fractions exceed the AGN fractions in the lower-density field environment by a mean factor of $6.1^{+10.3}_{-3.6}$. 

\citet{digby10} also measured the AGN fraction in Lyman-$\alpha$ emitters in their study of the $z=2.3$ protocluster HS 1700, as well as used the BX/MD color selection methods \citep{steidel03,adelberger04}. They report an AGN fraction of $2.9^{+2.9}_{-1.6}$\% for the Lyman-$\alpha$ emitters and $6.9^{+9.2}_{-4.4}$\% for the BX/MD sample. While their two samples are selected in different ways from ours, their X-ray luminosity threshold is similar. They do find that the fraction of Lyman-$\alpha$ emitters is similar to the field environment, which is similar to our result, but the AGN fraction in the BX/MD sample was approximately a factor of 20 higher. 

Finally, \citet{krishnan17} studied the $z=1.6$ protocluster Cl 0218.3-0510 and found that $17^{+8.6}_{-6.0}$\% of massive galaxies in the protocluster are AGN, whereas only $8^{+1.4}_{-1.2}$\% of similar field galaxies are AGN, or an excess of $2.1$ with $1.6\sigma$ significance. Both the cluster and field galaxy populations in this study were identified with multi-band photometry, rather than narrowband imaging, and therefore the galaxy and AGN selection is quite different from most of the other high-redshift protocluster studies. Figure~\ref{fig:sum} shows the evolution of the AGN fraction for high and low density regions, and Table~\ref{tbl:sum} summarizes our measurements and high-redshift measurements from the literature. 

\section{Summary} 

We have searched for AGN associated with the $z=2.53$ protocluster \uss\ with a deep, 100ks \chandra\ observation combined with follow-up spectroscopy with the MODS instrument on the LBT. This protocluster was discovered with deep, narrowband imaging around the $z=2.53$ radio galaxy 4C-00.62 by \citet{kajisawa06} and \citet{kodama07}, and a number of follow-up studies have established that it is one of the most substantial overdensities at $z>2$ \citep{hayashi12,hayashi16,shimakawa18}. 

While the \chandra\ data identified many X-ray sources in projection toward the protocluster, and unsurprisingly identified 4C-00.62 as an X-ray source, our follow-up spectroscopy did not identify any new AGN at the protocluster redshift. We did identify one X-ray AGN as a counterpart to one of the 100 HAEs identified by \citet{hayashi16} in the protocluster. We consequently conclude that the protocluster AGN fraction is $2.0^{+2.6}_{-1.3}$\% for X-ray AGN with $L_X > 2 \times 10^{43}$ erg s$^{-1}$ in HAEs with $L_{H\alpha} > 5.8 \times 10^{41}$ erg s$^{-1}$. If we remove the previously-known radio galaxy 4C-00.62 from this calculation, the fraction is $1.0^{+2.3}_{-0.8}$\%

The AGN fraction in \uss\ is lower than measurements of the AGN fraction in emission-line galaxies in two other overdensities at $z>2$. \citet{lehmer09} identified AGN in $5.1^{+6.8}_{-3.3}$\% in the Lyman-$\alpha$ emitters in the $z=3.09$ protocluster in SSA22 and \citet{lehmer13} identified AGN in $17^{+16}_{-9}$\% of HAEs in the $z=2.23$ protocluster 2QZ. However, \citet{digby10} identified AGN in only $2.9^{+2.9}_{-1.6}$\% of the Lyman-$\alpha$ emitters associated with the $z=2.3$ protocluster HS 1700, which is consistent with our measurement. 

Another notable aspect of these measurements is that our protocluster AGN fraction for HAEs is comparable to the AGN fraction for field HAEs. This is in contrast to most other studies that have found a significantly higher AGN fraction in protoclusters compared to field galaxies, and includes populations selected by emission lines and colors. While \citet{digby10} did not find an excess of AGN in their Lyman-$\alpha$ emitter sample, \citet{lehmer09} measured a factor of $6^{+10.3}_{-3.6}$ enhancement and \citet{lehmer13} measured a factor of $3.5^{+3.8}_{-2.2}$ enhancement. 

Our expectation was that we would measure a higher AGN fraction in \uss\ than in a similarly-selected field sample. This expectation was motivated by both previous studies of AGN in other protoclusters, and that the higher galaxy density in protoclusters should foster a greater rate of interactions and thus black hole accretion. The lower AGN fraction that we observe may simply represent a random fluctuation in the number of AGN, as the AGN fraction in protoclusters is based on a very small number of AGN and protoclusters, and most rely on quite incomplete spectroscopic surveys of plausible protocluster members. Another possibility is that \uss\ may be in a sufficiently early evolutionary stage that there are fewer sufficiently massive galaxies to host AGN. There is some evidence for this from the relatively small fraction of red galaxies, and the distribution of the known members into many subclumps \citep{kodama07,galametz12}. Finally, protoclusters likely span a wide range of total mass at their observed epoch, will be in a wide range of evolutionary stages, and they will likely evolve to clusters with a wide range of total mass by the present day. There could plausibly be trends between AGN fraction and protocluster mass and/or evolutionary state, but the current data for protoclusters is insufficient to address them. The best avenue for future progress is deep \chandra\ observations for other protoclusters, and most likely protoclusters identified as substantial overdensities of emission-line galaxies. This is because narrowband selection, combined with multi-color photometry, is the most efficient way to identify large numbers of galaxies at the same redshift. 

\section*{Acknowledgements}

We are grateful to Scott Adams and Kevin Croxall for helpful discussions about the analysis of MODS prism data. Support for this work was provided by the National Aeronautics and Space Administration through Chandra Award Number G04-15134X issued by the Chandra X-ray Observatory Center, which is operated by the Smithsonian Astrophysical Observatory for and on behalf of the National Aeronautics Space Administration under contract NAS8-03060. The work of PM is partially supported by the National Science Foundation under Grant 1615553 and by the Department of Energy under Grant DE-SC0015525.


\end{document}